\documentstyle[12pt]{article}

\newcommand{\newsection}{    
\setcounter{equation}{0}
\section}
\def\eop{\vspace*{\fill}\pagebreak}

\begin{document}

\begin{titlepage}
{\bf September, 1992}\hfill	  {\bf PUPT-1343}\\
\begin{center}

{\bf BOSE CONDENSATION AND $Z_N$ SYMMETRY BREAKING \\
IN THE MIXED MODEL OF INDUCED QCD}

\vspace{1.5cm}

{\bf  A.A.~Migdal}

\vspace{1.0cm}

{\it  Physics Department, Princeton University,\\
Jadwin Hall, Princeton, NJ 08544-1000.\\
E-mail: migdal@acm.princeton.edu}

\vspace{1.9cm}
\end{center}

\abstract{
The mixed model of the large $ N $ induced QCD, with $ N_f \ll N $
flavors of heavy fermions in fundamental representation, is solved  in
the local limit. The $ Z_N$ symmetry is broken spontaneously in the
large $ N $ limit, evading  the Elitzur "no-go" theorem. As a result
of this symmetry breaking, there is the Bose condensate of the
eigenvalues of the scalar field, proportional to $ \frac{N_f}{N} $.
This condensate leads to the  mass unit, which goes to zero as
fractional power of $ \frac{N_f}{N} $, thus defining the new kind of
the local limit of this lattice theory. There is a strong coupling
region below this mass scale, which revives the hopes of induction of
realistic QCD.
}
\vfill
\end{titlepage}


\newsection{Introduction}

In the previous paper~\cite{Mig92d} we suggested the mixed model of
induced QCD, with the local $Z_N$ symmetry broken by the extra spinor matter
in the fundamental representation of the $SU(N)$ gauge group. In the
old model~\cite{KM92,Mig92a,Mig92b,Mig92c}, built around the scalar field
in adjoint representation of the gauge group, this symmetry led to the
infinite string tension~\cite{KSW92,KhM92}, which was the main
obstacle for the induction of QCD.

Let us discuss some unusual features of induced theories, before going
into technical details.

It is not difficult to give an example of the model of induced QCD
with broken $Z_N$ symmetry and correct perturbative behavior at
intermediate scales.\footnote{There are {\bf two} strong coupling regions in
induced models, one at the lattice scales, and another at physical
hadron scales, with the perturbative region in between. In the local
limit, the whole range of scales expands, so that the lattice
"confinement region" becomes unobservable.} Just take large number
$N_f \gg N$ of flavors of the fermionic matter $ \Psi_x^i, i=1,\dots N_f $  in
the fundamental representation.
Then, the effective QCD Lagrangean $ \mbox{ tr } F_{\mu\nu}^2$ would
be induced with the factor $ N_f $, which corresponds to small
effective gauge coupling $ N g^2 \sim \frac{N}{N_f}$ in the
intermediate region.

Unfortunately, such model would not be solvable. There would be about
$ N_f^2 $ gauge invariant composite fields $ \bar{\Psi}_x^i \Psi_x^j $ at
each cite $x$ of the lattice after integrating out the gauge field. The
effective Lagrangean for these fields would be only about $ N^2 $ ( which
is the number of the gauge degrees of freedom integrated out). So, the
effective theory would be strongly coupled, even stronger than induced
QCD. For the solvable model, this should be the other way around. The
number of flavors should be much less, than the number of colors, so
that the effective theory for the composite fields would be weakly
coupled.

On the other hand, we want the induced QCD coupling to scale as $N^0
$, which requires about $N^2$ matter degrees of freedom. We could
achieve this by adding some field in the adjoint representation. The
unique property of the scalar field $ \Phi_x $ with one flavor is that
most of its $N^2 $ degrees of freedom can be absorbed into the gauge
field. Only $ N $ eigenvalues of this matrix are left, the rest can be
gauged away. One can go one step further, by introducing the
eigenvalues density $ \rho_x(\phi) $ as collective field, by local change
of variables~\cite{Mig92b}. Then the effective Lagrangean is $N^2 $
times greater than the number of the local degrees of freedom, so that
the WKB expansion could be developed around some classical solution for
$ \rho $. So, the model could be solved for arbitrary dimension $ D $
of the lattice.

Does this mean, that the model is trivial? Not necessarily. The
classical solution for this effective field theory involves the
quantum effects of original theory, since the majority of degrees of
freedom is integrated out. Should we instead integrate out the matter
field we would still get the induced gauge theory with correct
counting of $N$.

The question is, whether we are dealing with the
correct phase of the gauge theory, in other words, are we able to
reproduce the perturbative QCD at intermediate scales. This question
is a dynamical one, there are no kinematical restrictions, which would
immediately lead to the negative answer.  It is crucial though, that
there is the self-interaction of the scalar field. Without this
interaction one gets only the lattice artifacts, as it was observed in
numerical simulations~\cite{GS92}, as well, as in the analysis of the
exact solution~\cite{Gr92}.

With the self-interaction present, there is the following scenario of
induced QCD~\cite{KM92,Mig92a}. Let us assume we induced QCD and let
us go backwards in spatial scales, from the large ones to the lattice spacing.
The confinement and perturbative QCD regions are the same as usual,
all the way to the scale when the scalar field revives. When we go
from this region deeper into the ultraviolet direction, the effective
scalar coupling grows, and destroys the asymptotic
freedom.\footnote{For this end, the screening of the gauge
coupling by the interacting scalar field must overcome the usual
antiscreening. In absence of scalar interaction, the screening is like
$ 20 $ times smaller, than the antiscreening, as it is well known.
Presumably, the strong attraction is needed, to put together $ 20 $
times more scalars. The attraction of scalars in the
matrix models is a well-known phenomenon. Within the $ \frac{1}{ N } $
expansion it does not lead to instabilities, but beyond that it
requires nonperturbative stabilization by higher terms.}. Some
adjustment of parameters is needed, of course, otherwise we would not
reach the exactly infinite gauge coupling at the lattice scales. Also,
the bare scalar mass should be adjusted, otherwise the scalars would
decouple at the lattice scales, rather than in the perturbative
region.

The density of eigenvalues of the scalar field reflects the evolution
of the effective coupling as follows. The region of small (large) density
correspond to strong (weak) effective coupling of the effective field theory,
because the coupling is proportional to the ratio of the fluctuative
part to the classical one (or, to put it another way, the ordered
eigenvalues fluctuate only between neighbors, so, the denser they are
spaced, the less they fluctuate).

Since the scalar field has positive dimension of mass at $ D >2 $, the
scales of the eigenvalues are directly related to the mass scale. The
induced QCD scenario would correspond to small density $ \rho(\phi)
\rightarrow 0 $ at small and large $ \phi $, with the large density
region in between. The large $ \phi $ strong coupling region is
familiar. This is the string singularity, corresponding to the
endpoint of the support of the eigenvalues. The double scaling limit
of the matrix models of the string theory corresponds to the rescaling
of the vicinity of this endpoint at large $ N $.

The small $ \phi $ strong coupling region is new. It corresponds to
the infrared, rather than ultraviolet limit, where there are some
massless excitations present. This corresponds to the confinement
region in the induced QCD scenario. As for the large density in
between, this must be the region of the perturbative QCD, if all goes
according to the plan.

The exact solution of the model, for arbitrary potential, to some
extent agrees with this scenario. There are infinitely many fixed
points, which correspond to different power laws $ \rho(\phi)
\rightarrow |\phi|^{\alpha}, \cos \pi \alpha = -\frac{D}{3D-2}$  for
the density at the origin. The different values
of $ \alpha $ correspond to the different relevant terms in the scalar
potential. Namely, the term $ \Phi^{2m+2} $ contributes $ \phi^{2m+1}
$ to the classical equation for density, which is relevant for $
\alpha > 2m+1$. Thus, for $ 1<\alpha <3 $ only the mass term is
relevant, so that the mass should be adjusted. The quartic coupling
shows up in the next level solution, with $3 < \alpha < 5$, etc.

The simple dimensional counting, with  $ \Phi
\sim r^{1-\frac{1}{2} D} $, shows, that the $ \Phi^{2m+2} $ terms are
relevant in $ D \le 2 +  \frac{2}{m} $ dimensions. One, therefore, may
speculate, that the fixed point with $ 2m+1 < \alpha < 2m+3 $ is
realized at $ 2+ \frac{2}{m+1}< D \le 2 + \frac{2}{m} $ dimensions.
Thus, at $ 4 < D \le \infty $ the fixed point with $ 1 < \alpha < 3 $ is
realized, at  $ 3 < D \le 4 $ the one with $ 3 < \alpha < 5 $, etc.
At $ D \rightarrow 2 +0 $ the power of the potential grows, so that
it becomes nonpolynomial at $ D=2 $.

Formally, these solutions for the density apply for $ D=1 $ as well,
though, of course, the ultraviolet and the infrared regions are
interchanged here, as the scalar  field has the negative mass dimension.
There, the gauge field can be gauged away, so that the model reduces
to the $ C=1 $ string theory with the well-known exact solution: $
\rho(\phi) =  \frac{1}{ \pi} \sqrt{2( E - U(\phi) )} $.
In our solution, $ \alpha = 2k+1$, which corresponds to the upside-down
potential $ U_0 -\Phi^{4k+2} $, with the chemical potential at the
tip: $ E = U_0$. \footnote{At $ D \rightarrow 1 + 0 $ one could test the
solution, by perturbing the well-known free fermion Hamiltonian by the
spatial interaction with the gauge field. This would be an interesting
exercise, which might shed some light on the mysterious mechanism of
the induction of the gauge theory.}

The  physical mass spectrum, corresponding to the plane wave fluctuations of
the
vacuum density, also satisfies the scaling law $ m^2_{phys}  \propto
\epsilon^{\delta}$, where $ \epsilon  $ is the deviation of the bare
parameters from the critical values, and $ \delta = \frac{1}{2} +
\frac{k-2m-1}{2(\alpha -k)}$. The details can be found
in~\cite{Mig92b,Mig92c}.

These arguments applied to the  scalar model. In the mixed
model, with the perturbations, coming from the fundamental $ \Psi_x^{i} $
fields, the following new phenomena take place. These fields are
locally confined in the effective collective field theory, which
emerges after integrating out the gauge fields. In addition to the old scalar
density $ \rho_x(\phi) $, there is a flavor matrix density $
\sigma_x^{ij}(\phi)
$, which describes the $ \Psi $ condensate $ \sigma_x(\lambda) =
\bar{\Psi}_x\delta(\lambda-\Phi_x) \Psi_x $ with the color indexes
summed, but the rest of indexes fixed. There are simply no objects in this
collective field theory with  $ \Psi $ and $ \bar{\Psi} $ at different
points, in virtue of the  gauge invariance. In terms of Wilson
loops this local confinement corresponds to the  zero area law $ W(C)
= \delta_{0,A(C)} $.

Clearly, these $ \Psi $ fields cannot be quarks, they play different
role. Namely, they break the local $ Z_N $ invariance of the scalar
model. In the limit of $ N_f \ll N $, where the mixed model is
solvable, this breaking is proportional to $ \frac{N_f}{N} $, unless
there is a dynamical
enchancement. As we noted in~\cite{Mig92c}, such enhancement could,
indeed, take place. The $Z_N$ breaking combination of these
densities satisfies the linear homogeneous integral equation, with the
kernel depending on the old density. Under certain conditions, which
generically involve the extra parameter adjustment, this linear
integral equation would have finite solution. This is just the
symmetry breaking we are looking for. In the limit of $ N_f \ll N $,
when the model is solved classically, this breaking is spontaneous.

There is the well known "no-go" theorem due to Elitzur, which
forbids the spontaneous breaking of any {\em finite} local symmetry.
The point is, for the finite symmetry group one could trust the local
group integrations, which
eliminate the nontrivial representations from the set of local
observables. However, we are dealing with exceptional case, when these
group integrals diverge in the large $N$ limit. In fact, all our
critical phenomena are hidden in the Itzykson-Zuber integral over
unitary group. Much in the same way as in the Eguchi-Kawai model, the
spatial divergencies are imitated by the divergencies of these group
integrals. This is the key to understanding  the large $ N $ induced
models.

When the physical, light quarks are added to the mixed model, they
already can propagate in the vacuum. The classical Wilson loop would
now obey the perimeter law, since the global $Z_N $ symmetry is also
broken. Roughly speaking, the light quark screens its $Z_N$ charge by
picking up $ \bar{\Psi} $ from the vacuum, plus some number of $
\bar{\Psi} \Psi $ pairs. The decrement of the perimeter law could be
very large, though, if the symmetry breaking is small. So, the Wilson
loops in the local limit would satisfy the area law, if the string
tension would stay less than the mass cutoff, so that the area term
would overcome the perimeter term.

It is still unclear, how could one get the area law in the mixed model.
Clearly, it is impossible at the classical level of effective field
theory of two densities, so we need  the quantum effects
to be enchanced at large distances.

In this paper we are studying the mixed model in detail.

In the next Section 2. we reproduce the basic equations of the mixed
model~\cite{Mig92d}, with some generalizations made and some bugs
removed.

In the next Section 3. we go to the local limit, in the same way, as
in the scalar model, and find exact solution for the fermion
density. In the scalar density, we find a Bose condensation of the
eigenvalues at the origin, leading to the nonperturbative phenomena in
the infrared region.

In the Section 4. we study the wave equation for the scalar branch of
the spectrum. The Bose condensation leads to new mass unit, which
scales as a fractional power of $ \frac{N_f}{N} \rightarrow 0 $. We
argue, that the mass spectrum is infinitely raising, as fractional power of the
principal quantum number.

In the Section 5. we discuss the remaining problems, and suggest the
flavor instanton mechanism of the area law.

In the Appendix we compute the lowest scalar masses and the WKB
asymptotics of the higher ones in the mixed model.

\newsection{Basic equations of the mixed model}

The functional integral of the mixed model reads
\begin{eqnarray}
&&\prod_{x} \int D \Phi_x \exp - N \mbox{ tr }\left[ U(\Phi_x) \right]
 \int D \Psi_x \exp  \mbox{ tr }\left[ M \left(\Phi_x \right )\Psi_x
\bar{\Psi}_x \right]  \\
\nonumber & \,&
\prod_{<xy>} \int D \Omega_{xy} \exp N \mbox{ tr } \left[
\Phi_x\Omega_{xy}\Phi_y\Omega_{xy}^{\dag} +
\Omega_{xy} \Psi_y H_{xy}\bar{\Psi}_x +
 \Omega_{xy}^{\dag}\Psi_x H_{yx} \bar{\Psi}_y \right],
\end{eqnarray}
where  $ \Psi_x $($\bar{\Psi}_x$) is $ N \otimes N'
$($N'\otimes N$) matrix.
The second dimension $ N' =N_s\, N_f$ where $N_f$ is the number of
flavors, and $ N_s =  2^{\lfloor \frac{1}{2} D  \rfloor} $
is the number of spin components. The matrices $ H_{x, x+\mu} =
\frac{1}{2} \left(1+\gamma_{\mu} \right) $ act on the spin  components.

As compared to the previous paper~\cite{Mig92d}, we introduced the
direct interaction $ \bar{\Psi} M(\Phi) \Psi $ between the scalar and
spinor matter, instead of just the mass term $ \bar{\Psi} M \Psi $. This
does not spoil the solvability of the model. We also removed an
unnecessary factor of $N$ from this term.

The one-link integral
\begin{equation}
  I =\int d\Omega \exp N \mbox{ tr }\left[ \Phi_1 \Omega \Phi_2
\Omega^{\dag} + \Omega \Psi_2 H_{12} \bar{\Psi}_1 + \Omega^{\dag}
\Psi_1 H_{21}\bar{\Psi}_2  \right],
\end{equation}
depend, in virtue of the gauge invariance, only upon the densities
$ \rho_1,\rho_2, \sigma_1,\sigma_2 $, where
\begin{equation}
  \rho_x(\lambda) = \frac{1}{N } \mbox{ tr } \delta(\lambda-\Phi_x),
\label{RO}
\end{equation}
\begin{equation}
  \left \langle i \alpha|\sigma_x(\lambda)|j \beta \right \rangle  =
  \frac{1}{N} \mbox{ tr } \delta(\lambda-\Phi_x) \Psi_x^{i,\alpha}
\bar{\Psi}_x^{j,\beta}.
\label{SI}
\end{equation}
with the spin indexes $ \alpha, \beta $ and flavor indexes $ i,j $
fixed. We shall treat $ \sigma $ as $ N'\otimes N' $ matrix and use
the notation $ \mbox{ tr}' $ for the corresponding trace.

Changing the variables from $ \Phi_x, \Psi_x $ to these local
densities, we obtain the extra term in effective Lagrangean
\begin{equation}
  \delta S_{eff}(\sigma,\rho) =  N\, \int d \lambda
\rho(\lambda) \left(- N' \ln \rho(\lambda) +  \mbox{ tr}' \ln
\sigma(\lambda)  \right).
\end{equation}

As a result, we get the following classical equations for these two densities
\begin{equation}
 \frac{2D}{N^2}  \frac{d}{d \lambda} \frac{\delta \ln I}{\delta
\rho(\lambda)} + 2 \Re  V'(\lambda) =  U'(\lambda) + \frac{1}{N} \left(
- N'\,\frac{\rho'(\lambda)}{\rho(\lambda)} + \mbox{ tr}'
\frac{\sigma'(\lambda)}{\sigma(\lambda)}\right),
\label{EQRO}
\end{equation}
\begin{equation}
 \eta(\lambda) \equiv
\frac{2D\sigma(\lambda)}{N \rho(\lambda)}\frac{\delta \ln I}{\delta
\sigma(\lambda)} =
1- M(\lambda) \frac{ \sigma(\lambda)}{ \rho(\lambda)}
\label{EQSI}
\end{equation}
(The factor $ \rho(\lambda) $ in denominator of $ \eta $ was
unadvertedly skipped  in the previous paper~\cite{Mig92d}.)

Next problem is how to compute the variations of the one-link
integral $ I $  in these equations. In the same way, as in the scalar
model, the Schwinger-Dyson identities are used to derive the integral
equations for these two variations.\footnote{We assumed that the
classical solutions for $ \eta(\lambda), \sigma(\lambda) $ are
proportional to the unit $ N' \otimes N' $ matrix in virtue of symmetry.}
The first Schwinger-Dyson equation for the scalar density is the same
as in the scalar model
\begin{equation}
  \wp \int \, d \mu \left( \frac{\pi \rho(\mu)}{\mu-\lambda} + \arctan
\frac{\pi \rho(\mu)}{\lambda-R(\mu)} \right) =0,
\label{SD1}
\end{equation}
but with the extra terms in the effective potential
\begin{equation}
  R(\nu) = \frac{U'(\nu)}{2D}
 + \frac{N_f\,N_s}{2N\,D} \left(-\frac{\rho'(\nu)}{\rho(\nu)}
+ \frac{\sigma'(\nu)}{\sigma(\nu)} \right)
 + \frac{D-1}{D} \Re  V'(\nu + \imath 0)
\label{RLA}
\end{equation}
where
\begin{equation}
  V'(z) = \int  d \mu \frac{\rho(\mu)}{z-\mu} .
\end{equation}

The second Schwinger-Dyson equation reads
\begin{equation}
  \wp \int d\nu \frac{\rho(\nu)\eta(\nu)}{\lambda-\nu}
=\wp \int d\nu \frac{1}{ G^0_{\lambda}(\nu)} \,
\frac{(\lambda-R(\nu))}{(\lambda-R(\nu))^2 + \pi^2 \rho^2(\nu)}\,
\wp \int \,\frac{d\mu}{\pi^2} \frac{\eta(\mu)}{\nu-\mu},
\label{SD2}
\end{equation}
where
\begin{equation}
G^0_{\lambda}(\nu)= \frac{\Re  {\cal T}^0_{\lambda}(\nu +\imath 0
)}{\lambda-R(\nu)}\\;\;
{\cal T}^0_{\lambda}(z) = \exp \left( \int \frac{d \nu}{ \pi (\nu-
z )} \arctan \frac{\pi \rho(\nu)}{\lambda-R(\nu)}\right) .
\label{Solution}
\end{equation}
The Schwinger-Dyson equations also yield the auxiliary condition
\begin{equation}
  0=\int d\nu\eta(\nu).
\end{equation}
These equations are exact in the large $ N $ limit, for arbitrary
potentials $ U(\lambda) , M(\lambda) $.

\newsection{Bose condensation}

For any particular lattice model, there would be a
finite support of the eigenvalues, with the endpoint at the
lattice cutoff scale. In the local limit, we are going to ignore these
lattice artifacts, and extend the support to the whole real axis. The
ultraviolet divergencies would be dimensionally regularized. As a
result, there would be simple dispersion relations between the real and
imaginary parts of the functions, such as $ V'(z) $, which are
analytic in the upper half plane, with the symmetry
\begin{equation}
  V'(-\bar{z}) = - \bar{V}'(z).
\end{equation}
These dispersion relations read
\begin{equation}
  \Re  V'(\nu+ \imath 0 ) = \wp \int \, \frac{d\mu
\rho(\mu)}{\nu-\mu}\\;\;
\pi^2 \rho(\mu)= \wp \int \, \frac{d\nu \Re  V'(\nu+ \imath 0)}{\nu-\mu}.
\end{equation}

Also, note, that in virtue of the
symmetry of the scalar model, the density $ \sigma(\lambda) $ is an
even function of $ \lambda  $ as well, as the old density $
\rho(\lambda) $. Therefore, $ \eta(\lambda) $ is even as well, so that
the principle value integrals in (\ref{SD2}) are the odd functions.
The product of remaining factors in the $ \nu $ integral on the right of
(\ref{SD2}) are neither odd nor even; hence, only the odd part of this
product can be left.

With this property in mind, we can expand this product in powers of $ \rho $
and $ r(\lambda) \equiv R(\lambda) - \lambda $  in
the local limit, in the same way, as in the previous works.
The first odd term would be a linear one, and we find
\begin{equation}
  \wp \int d\nu \frac{\rho(\nu)\eta(\nu)}{\lambda-\nu}
= \wp \int \, d \nu \int \frac{d \phi
\rho(\phi)}{(\nu-\phi)(\lambda-\phi)}\,
\wp \int \,\frac{d\mu}{\pi^2} \frac{\eta(\mu)}{\nu-\mu},
\end{equation}

Integrating over $ \phi $ and $ \nu $ on the right side, and using
dispersion relations, we find a very simple result
\begin{equation}
  \wp \int \, \frac{d \mu \eta(\mu)}{\lambda-\mu} \left(\rho(\mu)
-\rho(\lambda) \right),
\end{equation}
which leads to the puzzling equation
\begin{equation}
  0 = \rho(\lambda) \wp \int \, \frac{d \mu \eta(\mu)}{\lambda - \mu}.
\label{ZET}
\end{equation}

How could it be satisfied for finite $ \eta $? Recall that all our
integrals are understood
in the sense of analytic continuation of the powerlike integrals. For
the powerlike Anzatz $ \eta(\mu) = \mu^{\gamma} $ we find
\begin{equation}
  0 = \rho(\lambda) \lambda^{\gamma}\tan \frac{ \pi \gamma }{2}
\end{equation}
This equation would hold for any even $ \gamma $. Positive values of $
\gamma $ correspond to nonsingular density $ \eta $ which is dominated by the
lattice scales. The singular solutions, with $ \gamma = - 2 n $ are
more interesting. With proper normalization this corresponds to
\begin{equation}
  \int d \mu \frac{\eta(\mu)}{\mu-z} \propto \imath  z^{-2n}\\;\;
 \wp \int \, d \mu \frac{\eta(\mu)}{\mu-\lambda} \propto \Im
\left(\lambda+ \imath 0 \right)^{-2n} \propto  \delta^{(2n-1)}(\lambda).
\end{equation}

Coming back to (\ref{ZET}), we see that this solution holds for the
density $ \rho(\lambda) $ which vanishes faster than $ \lambda^{2n-1}
$ at the origin. Note, that the auxiliary condition is
also satisfied for any {\bf positive integer} $ n $.

Now, let us substitute this solution into the first classical equation
and find the first $ \frac{N_f}{N} $ correction to  the scalar
density. Simple algebra yields, up to irrelevant regular terms, at $
\nu  \rightarrow 0 $
\begin{equation}
  \frac{N_f\,N_s}{2N\,D} \left(-\frac{\rho'(\nu)}{\rho(\nu)}
+ \frac{\sigma'(\nu)}{\sigma(\nu)} \right) \rightarrow
- \frac{n N_f N_s}{ND}  \frac{1}{\nu}
\end{equation}
This term can be compensated in $ R(\nu) $ by the last term, if the
following $ \delta $-term would be added to the scalar density
\begin{equation}
  \delta \rho(\lambda)= \frac{n N_fN_s}{(D-1) N} \delta(\lambda).
\end{equation}
This term is positive, as it should be. This is the Bose-condensation
of the eigenvalues.

Note, however, that the sign changes at $ D <  1 $. In this case, the
gap must develop, to avoid negative density. Or, to be more explicit,
the pole term in  $ R(\nu) $ at $ D < 1 $ can be compensated by the
extra term in the density
\begin{equation}
  \delta \rho(\lambda) = -\rho(\lambda) \Theta(b^2 - \lambda^2)
\end{equation}
which effectively cuts the dispersion integral at $ \pm b $. The
corresponding variation
\begin{equation}
 \frac{D-1}{D} \wp \int \, d \nu \frac{\delta \rho(\nu)}{\lambda -
\nu}  \rightarrow - \frac{2 (D-1)b^{\alpha+1}}{D(\alpha+1)}  \frac{1}{ \lambda}
\end{equation}
would cancel the pole term provided
\begin{equation}
  b^{\alpha+1} = \frac{(\alpha+1)n N_f N_s}{2(1-D) N}.
\end{equation}

This is an interesting phenomenon. The transition at $ D =1 $
corresponds to the interplay between the repulsion of the eigenvalues,
caused by the usual Vandermonde determinant, and the attraction,
caused by the gauge field via the Itzykson-Zuber determinant. Namely,
the increase in the density at the origin raises the Vandermonde term
in effective energy and lowers the Itzykson-Zuber term. At $ D > 1 $
the latter wins. As for the interaction with fermions, it effectively
raises the energy, due to the Fermi repulsion. The net classical
forces are balanced, therefore, by compensating this repulsion by the
Itzykson-Zuber attraction.

Let us now look at the local limit of the first classical
equation~\cite{Mig92a}
\begin{equation}
 2 r(\lambda) \rho(\lambda) = \wp \int \, d \mu
\frac{\rho^2(\mu)}{(\mu-\lambda)}.
\end{equation}
Perturbing $ \rho $ in this equation ($ r(\lambda) = R(\lambda)
-\lambda $ is not perturbed), we find in the first order
\begin{equation}
  r(\lambda) \delta \rho(\lambda) = \wp \int \, d \mu \frac{
\rho(\mu) \delta \rho(\mu)}{\mu -\lambda}.
\end{equation}

This equation is satisfied for the $ \delta \rho(\lambda) \propto
\delta(\lambda) $ because $ r(0)=\rho(0) =\rho'(0) =0 $. The vanishing first
derivative $ \rho'(0)= 0$ is also needed, otherwise the right side
would not vanish at $
\lambda = 0 $. Vanishing of higher derivatives of $ \rho(\lambda) $ at
the origin is not needed here.

The $ \delta $-function at the origin is probably just the
result of the $ \frac{N_f}{N} $ expansion. We expect this delta
function to be smeared for finite $ \frac{N_f}{N} $. Actually, the
shape of the  density in the region $ \lambda^{\alpha+1} \sim
\frac{N_f}{N} $ is beyond the perturbation theory. In this region, the
higher orders of the $ \frac{N_f}{N} $ expansion are important,  and
we expect some scaling function
\begin{equation}
  V'(z) = z^{\alpha}f \left(\frac{N_f}{N z^{\alpha+1}} \right), f(x) =
f_0 + f_1 x + \dots.
\end{equation}

In other words, there is an infrared catastrophe in this model, taking
place at small scalar fields, which is equivalent to small mass
scales. Below the "confinement scale"
\begin{equation}
  z_0 \sim \left(\frac{N_f}{N}\right)^{ \frac{1}{ \alpha+1} }
\end{equation}
the quantum effects are significant for arbitrarily small $
\frac{N_f}{N} $. Hopefully, the dominant terms are easier to sum up,
than the QCD planar graphs.

%

\newsection{Scalar wave equation}

Let us study the wave equation for the fluctuations of the scalar
density. The wave equation was derived in~\cite{Mig92b} and studied
in~\cite{Mig92c}. It reads (at $ z = \lambda + \imath 0 $)
\begin{equation}
 -\frac{1}{2} \left(P^2+ M^2_{eff}  \right)\,\Im {\cal F}=
D \,\Re {\cal P}' \,\Im {\cal F} - \Im {\cal P}\, \Re{\cal F}' +
\frac{(D-1)^2}{D} \Im  {\cal P}
\left(\frac{ \Re {\cal P}\, \Im  {\cal F}}{\Im  {\cal P}} \right)',
\end{equation}
where the analytic function
\begin{equation}
  {\cal P}(z)= \frac{\imath A}{\cos \frac{\pi \alpha}{2}}
 \left(-\imath z \right)^{\alpha},
\end{equation}
is the same as before.

$ P $ is the Euclidean 4-momentum of the vacuum excitation,
corresponding to plane wave fluctuations of $ \rho $,
\begin{equation}
\Delta \rho(\lambda,x) =  \frac{1}{ N} e^{\imath P x}
 \wp \int \, d\nu \frac{\Im {\cal F}(\nu+ \imath0)}{\pi^3\rho(\nu)}
  \frac{1}{\nu-\lambda}.
\end{equation}

%
The effective mass $ M_{eff}^2 $ gets the double pole term
\begin{equation}
  M^2_{eff}(z) \equiv \frac{ D U''(z)-2D^2 -2 D +2}{D-1} +
\frac{a^2}{z^2} =\frac{a^2}{z^2}+ \tau_0  + \dots,
\end{equation}
with
\begin{equation}
  a^2 = \frac{2D n N_f N_s}{N(D-1)},
\end{equation}
which is the net result of the Bose condensation. The higher terms of
the expansion are irrelevant, since the dominant $ z \sim
\frac{a}{\sqrt{\tau_0}} $ are small at $ N_f \ll N $.

The boundary condition at $ \lambda  \rightarrow 0 $ is that $
\rho(\lambda)\Im  {\cal F}(\lambda+\imath 0 )=0 $, to kill the $
\delta(\lambda) $ term, which comes from the factor $
\frac{\lambda}{|\lambda|} $ in the ratio of the real to imaginary
parts of $ {\cal P} $. This boundary condition allows for the strong
singularity  $ \Im  {\cal F}(z) \propto z^{\omega} , \omega > - \alpha
$. However, with the $ \delta$-like perturbation in $ \rho $, this is
replaced by a weaker condition $ \omega > 0 $.

Unfortunately, we cannot impose the boundary condition at $ z =0 $,
because of the abovementioned infrared catastrophe. All we can do at
this point is to impose some matching condition at $ z \sim z_0 $,
where the higher corrections to the density become important. We could
demand, that $ \Im  {\cal F} $ vanishes at $ z_0 $, or take any other
finite value. As we shall see shortly, the generic asymptotics for
$ {\cal F} $ is exponentially large, therefore, the actual matching
value of $ {\cal F} $ does not matter. Only the leading exponential
term should cancel.

Let us turn to the wave equation and consider  the infinite sum of
power terms as an Anzatz for $ {\cal F} $
\begin{equation}
  {\cal F} = \imath^s\sum_{\epsilon}\, \frac{f(\epsilon) \left(-\imath z
\right)^{\epsilon}  }
{\sin \frac{\pi (s-\epsilon)}{2}},
\end{equation}
where $ s= \{ 0,1\} $ is the "$ \lambda $-parity"
\begin{equation}
  {\cal F}(-\bar{z}) = (-1)^s \bar{ {\cal F}}(z) \\;\; \Delta
\rho(-\lambda,x) = (-1)^s \Delta \rho(\lambda,x).
\end{equation}
Differentiating real and imaginary parts of $ {\cal P} , {\cal F} $,
multiplying by $ \Im  {\cal P} $ and collecting  the power terms, we
find the following equation
\begin{equation}
  \sum_{\epsilon} f(\epsilon) \left(
\lambda^{\epsilon+\alpha-1 } A C(\epsilon) +
\lambda^{\epsilon}(P^2+ \tau_0)+
a^2\lambda^{\epsilon-2} \right) =0,
\end{equation}
where
\begin{equation}
  C(\epsilon)=2 \epsilon  \cot
\left(\frac{\pi (\epsilon -s) }{2} \right) +2 \left(
\frac{(D-1)^2}{D} \epsilon +\alpha  D\right) \tan\frac{\pi \alpha}{2},
\end{equation}

The corresponding recurrent relation reads
\begin{equation}
  f(\epsilon) = -\frac{(P^2+\tau_0)f(\epsilon+\alpha-1) +
a^2 f(\epsilon+\alpha+1)}{A C(\epsilon)}.
\end{equation}
The leading power $ \epsilon_0 $ satisfies the transcendental equation
\begin{equation}
  C(\epsilon_0) = 0,
\end{equation}
which has infinitely many real solutions, approaching the integers
at large $ \epsilon_0 $.

There are two scaling variables involved,
\begin{equation}
  {\cal F}(z) = \imath^s \left(-\imath z \right)^{\epsilon_0}
{\cal G}\left(\frac{P^2+\tau_0}{A\left(-\imath z \right)^{\alpha-1}},
\frac{a^2}{z^2(P^2+\tau_0)} \right),
\end{equation}
where the scaling function $ {\cal G}(u,v) $ expands in double series
\begin{equation}
  {\cal G}(u,v) = \sum_{n=0}^{\infty}\frac{u^n}
{\sin \left(\frac{\pi}{2} (s+(\alpha-1) n-\epsilon_0)\right)}\, \sum_{l=0}^{n}
v^l \varphi_{n}(l) .
\end{equation}

The expansion coefficients $ \varphi_n(l)$ satisfy recurrent relation
\begin{equation}
    \varphi_n(l) = -
\frac{\varphi_{n-1}(l)+\varphi_{n-1}(l-1)}{C\left(\epsilon_0-n(\alpha-1)-2l\right)}
\end{equation}
with initial condition
\begin{equation}
  \varphi_0(l) = \delta_{l,0}.
\end{equation}
which is suitable for numerical or analytical computations. One can
estimate, that $ \varphi_n(l) $ decreases as $  \frac{1}{ n!} $, which is
sufficient for the absolute convergence of the expansion.

The spectrum is supposed to be defined from the boundary conditions at
\begin{equation}
  z \sim z_0 \equiv \left( \frac{a^2}{A} \right)^{\frac{1}{\alpha+1}}.
\end{equation}
This would
yield the following mass spectrum
\begin{equation}
  -P^2 = M_L^2=\tau_0 + A \left(\frac{a^2}{A}
\right)^\zeta\,\xi_L(s,\epsilon_0).
\end{equation}
with the critical index
\begin{equation}
  \zeta = \frac{\alpha-1}{\alpha+1}, 0 < \zeta < 1.
\end{equation}
The scaling variables are finite, so that the expansion
can be used to find the roots numerically. The results are presented
in Appendix. They correspond to the lowest energy levels.

The higher levels can be found analytically.
In this limit the first variable tends to $\infty $, and the second one
tends to zero. Leaving the $ l=0 $ term , and using the WKB
asymptotics for $ \varphi_n(0) $, we find the following spectrum (see
Appendix  for the details)
\begin{equation}
 \xi_L(s,\epsilon_0)  \propto L^{\zeta}, L \rightarrow \infty
\end{equation}

\newsection{Discussion}

The mixed model looks much better than the original scalar one. The
Bose condensation, caused by interaction with the fermions, leads to
the  nonperturbative phenomena in the infrared region. There is the new
mass scale $ M $, which goes to zero as a fractional power of the ratio of
flavors to colors $ M^2 \propto \left(\frac{N_f}{N}\right)^{\zeta} $ .
Beyond this scale one cannot neglect higher terms of the $
\frac{N_f}{N} $ expansion.

There are two sources of these terms. Some terms come from the
nonlinearities of the classical equations for the density. Those are
easy to account. The difficult  terms come from the higher quantum
corrections, starting from $ \frac{N_f^2}{N^2} $. At $ 1 \ll
N_f \ll N $ only the  fluctuations of the fermion condensate are important,
those of
the scalar density are damped by the $  \frac{1}{ N } $ factors. It is
a new interesting problem to sum up all the relevant infrared terms;
at present we do not know how to do it.

It is instructive to compare this situation with the one at large $
\frac{N_f}{N} $ when we cannot solve the model, but know that it
induces QCD. In this limit the induced gauge coupling constant goes to
zero as $ \frac{N}{N_f} $, which leads to exponentially small physical
mass scale in the lattice units. The scalar field and its self
interaction is irrelevant in this limit.

When the ratio $ \frac{N_f}{N} $ decreases, this scalar interaction
becomes important. In absence of it, there would be a phase
transition, at $ \frac{N_f}{N} = \frac{11}{2} $ when the induced self
interaction of the hard gluons takes over the  screening of the
running gauge coupling by the matter. At smaller $ \frac{N_f}{N} $ this
coupling would run to zero in the ultraviolet region.

The induced QCD scenario corresponds to the situation, when  strong
enough scalar coupling prevents this
phase transition,  pushing it all the way to $ \frac{N_f}{N} = 0 $.
Then, at small $ \frac{N_f}{N} $ we would still be in the QCD phase.
This possibility would display itself as the infrared catastrophe. At
small enough mass scales the nonperturbative effects would become
important. If our scaling solutions correspond to this phenomenon, then
we could estimate, that effective gauge coupling goes to zero, as
inverse logarithm of $ \frac{N_f}{N} $
\begin{equation}
  N g_0^2 \propto  \frac{1}{- \ln M^2} \propto  \frac{1}{ \ln
\frac{N}{N_f} }
\end{equation}

The singular behavior of the effecfive gauge
coupling and the mass scale with $ \frac{N_f}{N} $ does not qiuite meet
the old expectations. From the point of view of the usual RG analysis,
this singularity could come about due to the new
fixed point. The small perturbation from the
fermions could drive the scalar coupling to this new fixed point, in
which case the correlation length would, indeed, scale as some
fractional power of $ \frac{N_f}{N} $. The physical meaning of this
mysterious phenomenon is yet to be understood.

How are the gluons reproduced in terms of the effective field theory
of  the gauge invariant local composite fields $
\sigma(\lambda), \rho(\lambda) $? Clearly, the classical solutions for $
\rho, \sigma $ do not describe the long range fluctuations of gauge
field. This must be either the instantons of this effective theory, or
the quantum fluctuations. At small $ \frac{N_f}{N} $ the quantum
fluctuations are damped, so, perhaps, the flavor instantons  could be
responsible for gluons.

Take the Wilson loop, for example. The space independent classical
solution would simply yield the product of classical values of the
link matrices, thus leading to perimeter law. As for the instanton, it
potentially could yield the area law, as we know from the variety of
models in lower dimensions. All it takes, is the distortion of the
flavor matrix field $ \sigma_x(\lambda) $ in the region of the topology of
the disk, bounded by the  loop. Say, for the boundary conditions with
the $ 2 \pi n $ rotation in one of the $ U(1) $ flavor subgroups along
the  loop, there must be instantons, with the classical action,
proportional to the area, as it follows from the usual homological
arguments.

Leaving aside the Wilson loop, and the stringy mesons it is supposed
to describe, let us turn to the Goldstone-like mesons, corresponding
to collective excitations in our vacuum densities $ \sigma, \rho $.
The $ \rho $ excitations were studied in this paper. The $ \sigma $
excitations could be studied by the same method. Unlike the
(simplest fixed points of the) scalar model~\cite{Mig92b,Mig92c}, the
mixed model has {\em infinite} mass spectrum, raising as fractional
power of the principal quantum number. This is a good sign, supporting
the conjecture of induction of QCD.

At this point it is too early to make definite conclusions. The
euphoria of the first months of playing with the new toy has passed,
but it would be equally silly to turn away from this model, just because it
is not quite QCD. In fact, this is a whole new class of solvable models, which
deserves serious study, as it teaches us nonperturbative Quantum Field
Theory. As a by-product, we might induce QCD, or pass the $ D=1 $
barrier in the string theory, or find something completely unexpected.

\section{Acknowledgements}

This work was partially supported by the National Science Foundation under
contract PHYS-90-21984.

\eop

\eop
\appendix

\newsection{Spectral equation}

We imposed the boundary condition $ \Im  {\cal F }=0 $ at a particular point $
z =
\frac{1}{3} z_0 $, and computed the spectrum numerically for $ D=4,
\alpha = 3.36901$. The Taylor expansion
for the logarithmic derivative $ \frac{\partial}{\partial P^2} \Im
\ln {\cal F}$
in powers of $ x = (P^2 + \tau_0) $ with $ 50 $ digits of
accuracy in coefficients was converted to the
$ (9,9) $ and $ (10,10) $ Pad\'e approximants, after which the roots of
denominators were found.
The accuracy $ \Delta x_i $ of the poles $ x_i $ of this
function was estimated as the difference between the roots  of
denominators of these two approximants.

Here are the values of $
\epsilon_0\mbox{:} \; \{\{x_1,\Delta x_1 \},\dots \} $  for
the first ten levels of $ \epsilon_0 $. Only real roots with $ \frac{|
\Delta x |}{|x|} < .01 $ are left. The empty list implies that no such
roots were found in our approximation.
\begin{eqnarray}
 \matrix{ 2.05 :& \{ \{ -53.3,-0.0279\} ,
   \{ -36.6,-0.0000478\} ,\{ 12.,1.59\,{{10}^{-11}}\} ,
   \{ 27.8,1.07\,{{10}^{-7}}\} \}  \cr 4.07 :&
  \{ \{ -21.5,-6.14\,{{10}^{-10}}\} ,
   \{ 31.8,2.4\,{{10}^{-7}}\} \}  \cr 6.09 :&
  \{ \{ 21.1,-1.29\,{{10}^{-9}}\} \}  \cr 8.11 :& \{ \}
   \cr 10.1 :& \{ \{ 23.2,5.86\,{{10}^{-9}}\} ,
   \{ 51.8,0.00266\} \}  \cr 12.1 :&
  \{ \{ 10.7,-0.000486\} \}  \cr 14.1 :&
  \{ \{ 2.83,-2.22\,{{10}^{-10}}\} \}  \cr 16.1 :&
  \{ \{ -5.11,2.66\,{{10}^{-8}}\} ,
   \{ -3.71,7.85\,{{10}^{-11}}\} ,
   \{ 8.66,-6.86\,{{10}^{-7}}\} ,\{ 17.9,-0.00165\} \}
   \cr 18.1 :& \{ \{ -2.54,-1.2\,{{10}^{-11}}\} ,
   \{ 3.41,-5.18\,{{10}^{-12}}\} ,
   \{ 20.5,-0.0000458\} \}  \cr 20.1 :& \{ \}  \cr  }.
\end{eqnarray}
The large roots cannot be computed by this method, as the convergence
slows down. The WKB method can be applied in this limit.
Let us set $  A=1 $, and neglect $ \tau_0 $ compared to $ P^2 $. The
asymptotic form of $ \varphi_n(l) = \phi  \left(\epsilon_0-(\alpha-1)n-2l
\right) $ follows from the recurrent equation
\begin{equation}
  \phi (\epsilon) \sim \Gamma
\left(\frac{C(\epsilon)}{(\alpha+1)C'(\epsilon)} \right)
\left(-(\alpha+1) C'(\epsilon) \right)^{\frac{\epsilon}{\alpha+1}}
\\;\; \epsilon = \epsilon_0 -(\alpha-1)n - 2l.
\end{equation}

The sum over $ l $ converges at  $ l  \ll n $, since the second
argument of the scaling function is small at large $ |P^2| $. The
remaining sum over $ n $ is dominated by the pair of two complex
conjugate saddle points in $ \epsilon $ plane,  given by equation
\begin{equation}
 \left( \ln \phi(\epsilon) \right)'= \ln \frac{P^{\frac{2}{\alpha-1}}}{z}
\end{equation}

At large complex $ \epsilon $
\begin{equation}
  C'(\epsilon) \rightarrow 2 \tan \left( \frac{\pi \alpha}{2} \right)
\frac{(D-1)^2}{D}- 2 \imath \mbox{ Sign} \left( \Im  \epsilon \right)
\end{equation}
which yield complex conjugate  constants  $ C', \bar{C'} $ for the two
saddle points $  \epsilon,\bar{\epsilon} $. As a result, we find
\begin{equation}
  \epsilon = -\frac{a^2}{e
C'}\left(\frac{P^2}{z^{\alpha-1}}\right)^{\frac{\alpha+1}{\alpha-1}}
\end{equation}

The logarithmic derivative can be estimated as follows
\begin{equation}
   \frac{\partial}{\partial P^2} \Im \ln {\cal F} = \left \langle n-l
\right \rangle  \rightarrow  \left \langle n \right \rangle
\rightarrow  \frac{\Re \left( \epsilon\, e^{\imath \Phi}
\right)}{(1-\alpha) \Re  \left(e^{\imath \Phi} \right) }=
\frac{1}{\alpha-1} \left( - \Re (\epsilon) + \tan (\Phi) \Im  (\epsilon)
\right)
\end{equation}
where $ \Phi $ is the phase of the integrand at the saddle point,
\begin{equation}
  \Phi = \Im \left(\ln \phi + \epsilon \ln z - \frac{\epsilon}{\alpha-1}
\ln P^2 \right)
\end{equation}

The spectrum is, therefore, quantized as follows,
\begin{equation}
  \Phi = \pi \left(L + \frac{1}{2} \right)
\end{equation}

After simple transformations we find
\begin{equation}
 \frac{a^2}{(\alpha+1)e} \,  \, \Im  \left(
\frac{-e^{\frac{2\pi \imath}{\alpha-1}}}{C'} \right) \,
\left(-\frac{P^2}{z^{\alpha-1}} \right)^{\frac{\alpha+1}{\alpha-1}}= \pi
\left(L+
\frac{1}{2} \right)
\end{equation}
which yields the spectral equation in the text.

\end{document}